\documentclass[runningheads]{llncs}
\usepackage{graphicx}
\usepackage{comment}
\usepackage{amsmath,amssymb} 
\usepackage{color}

\usepackage{tabularx}
\newcolumntype{C}[1]{>{\centering\arraybackslash}p{#1}}
\usepackage{multirow}
\usepackage{hhline}

\begin{document}
\pagestyle{headings}
\mainmatter
\def\ECCVSubNumber{100}  

\title{L2-Constrained RemNet for Camera Model Identification and Image Manipulation Detection} 

\titlerunning{L2-Constrained RemNet}

\author{Abdul Muntakim Rafi\inst{1} \and Jonathan Wu\inst{1} \and
Md. Kamrul Hasan\inst{2}}
\authorrunning{Rafi et al.}

\institute{University of Windsor, 401 Sunset Ave, Windsor, ON N9B 3P4, Canada\and
Bangladesh University of Engineering and Technology, Dhaka-1205, Bangladesh\\
\email{\{rafi11,jwu\}@uwindsor.ca, khasan@eee.buet.ac.bd}}
\maketitle

\begin{abstract}		
		
Source camera model identification (CMI) and image manipulation detection are of paramount importance in image forensics. In this paper, we propose an L2-constrained Remnant Convolutional Neural Network (L2-constrained RemNet) for performing these two crucial tasks. The proposed network architecture consists of a dynamic preprocessor block and a classification block. An L2 loss is applied to the output of the preprocessor block, and categorical crossentropy loss is calculated based on the output of the classification block. The whole network is trained in an end-to-end manner by minimizing the total loss, which is a combination of the L2 loss and the categorical crossentropy loss. Aided by the L2 loss, the data-adaptive preprocessor learns to suppress the unnecessary image contents and assists the classification block in extracting robust image forensics features. We train and test the network on the Dresden database and achieve an overall accuracy of 98.15\%, where all the test images are from devices and scenes not used during training to replicate practical applications. The network also outperforms other state-of-the-art CNNs even when the images are manipulated. Furthermore, we attain an overall accuracy of 99.68\% in image manipulation detection, which implies that it can be used as a general-purpose network for image forensic tasks. 

\end{abstract}

\section{Introduction}

Camera model identification (CMI) and image manipulation detection are crucial tasks in image forensics with applications in criminal investigations, authenticating evidence, detecting forgery, etc. Digital images go through various camera-internal processing before being saved in the device \cite{kirchner2015forensic}. Moreover, they are often manipulated after they leave the device that has been used to capture them. Nowadays, professional image editing tools like Adobe Photoshop, ACDsee, and Hornil Stylepix are readily available, consequently making image manipulation a common phenomenon \cite{chen2019multi}. Also, images undergo different kinds of manipulations when they are shared online. We have observed a proliferation of digitally altered images with the advent of modern technologies. When the authenticity of such images is questioned, a forensic analyst has to answer two questions first, what is the source of the image under question and whether the image has been manipulated. The image metadata cannot be trusted as a reliable source, as this data can be forged. Therefore, a forensic analyst resorts to different image forensics techniques to answer these questions. 

Image forensics is an active research area, and several methods exist in the literature for finding out the source camera model and detecting image-processing operations of a questioned image. But researches are conducted discretely for finding out the source and manipulation history of an image. In \cite{stamm2013information}, \cite{piva2013overview}, we can find a brief overview of the approaches proposed over the last two decades. We see that initial research in CMI has focused on merging image-markers, such as watermarks, device-specific code, etc. \cite{piva2013overview}. However, using separate external features for each camera model is an unmanageable task \cite{farid2009image}. Consequently, researchers have focused on utilizing the intrinsic features, such as the Color Filter Array (CFA) pattern \cite{bayram2005source}, interpolation algorithms \cite{kharrazi2004blind}, and Image Quality Metrics (IQM) \cite{gloe2012feature}. Utilizing Photo Response Non-Uniformity (PRNU) noise patterns have been proposed for device-level identification \cite{dirik2007source}, \cite{fridrich2006digital}. Although sensor noise carries device-specific noise artifacts, researchers have developed methods to perform CMI using sensor noise patterns \cite{thai2014camera}, \cite{lukas2006digital}. Most of these approaches attempt to extract camera model-specific features and compare the features with a pre-calculated reference for the corresponding camera model \cite{cao2009accurate}. In the case of image manipulation, traces are found in the image according to the type of processing it has gone through \cite{bayar2018constrained}. Following this theory, researchers have used distinct forensic approaches for identifying different kinds of image manipulation, such as resizing \cite{popescu2005exposing}, \cite{feng2012normalized}, contrast enhancement \cite{stamm2010forensic}, \cite{yao2009detect}, and multiple jpeg compression \cite{bianchi2012image}, \cite{neelamani2006jpeg}, etc. The drawback of using the above-mentioned statistical feature-based approaches is that the performance degrades sharply, when new cases arise that have not been considered during feature vector selection \cite{chen2019multi}. For that reason, more recent researches have focused on becoming data-driven, such as utilizing local pixel dependencies used in steganalysis \cite{fridrich2012rich}, \cite{pevny2010steganalysis} to perform CMI \cite{marra2017study}, \cite{chen2015camera} and detect image manipulation \cite{qiu2014universal}. In \cite{fan2015general}, the authors propose a Gaussian mixture model for image manipulation detection. Though these approaches provide good results, extracting features for different manipulations requires substantial computational resources, and the performance degrades severely depending on the size of the questioned image \cite{chen2019multi}.

Recently, researchers have started applying Convolutional Neural Networks (CNNs) for image forensic tasks \cite{yang2020survey}. It is expected as CNNs have performed extremely well in different image classification tasks \cite{schmidhuber2015deep}. Usually, CNNs tend to learn features related to the content of an image, whereas, for image forensics, we need to refrain CNNs from learning image contents \cite{bayar2017design}. As a result, a common practice while using CNNs in digital image forensics is adding a preprocessing layer at the beginning of the CNN architecture. Chen et al. \cite{chen2015median} have proposed using a median filter, whereas Tuama et al. \cite{tuama2016camera} have used a high-pass filter before feeding images in their respective CNNs. However, such crude filtering is not supported by the literature as the artifacts introduced by different camera-internal processing and manipulations can lie in both low and high frequency domain \cite{lukas2006digital}. Therefore, fixed filters as preprocessor may lose forensics-related features. Bayar and Stamm \cite{bayar2017design} have proposed a data-driven constrained convolutional layer which has performed better than the above-mentioned fixed filters. Rafi et al. \cite{rafi2019remnet} have used a completely data-driven preprocessor block followed by a classification block to perform CMI. Bayar and Stamm \cite{bayar2017design} have also used their constrained CNN for image manipulation detection. However, some CNN based approaches do not use any preprocessing scheme. Yang et al. use the idea of multi-scale receptive fields on an input image to perform CMI \cite{yang2017source}. In \cite{bondi2017first}, the authors use CNN and support vector machine (SVM) for CMI, where they use the CNN part as a feature extractor. In \cite{rafi2019application}, explores the performance of DenseNet \cite{densenet} in both CMI and image manipulation detection. In \cite{chen2019multi}, the authors investigate the performance of densely connected CNNs in image manipulation detection. Owing to the performance of the data-driven preprocessing schemes, it can be inferred that further researches need to be conducted to make the preprocessing operations more robust for image forensic tasks. Several researches exist in the literature that use auxiliary loss function to enhance the discrimination between learned features \cite{wen2016discriminative}, \cite{wen2016latent}, \cite{sun2015deeply}. There is a scope of utilizing such auxiliary loss functions in the modular CNN architectures for image forensics.

Despite the numerous researches conducted in this field, most researchers have explored CMI and image manipulation detection problems discretely. Bayar and Stamm \cite{bayar2017design} show that it is possible to use the same approach for both tasks. Therefore, research for coming up with a general-purpose neural network suitable for both CMI and image manipulation detection requires more attention. Also, strict measures should be followed while conducting experiments so that the proposed methods can be applied in real-life scenarios. Kirchner and Gloe suggest that the test set should always consist of images captured by devices that have not been used during training or validation \cite{kirchner2015forensic}. Also, the scenes in the test set should be different from those used during training and validation. Here, \emph{scene} refers to a combination of a location and a specific viewpoint. Keeping separate devices and scenes in the test set is compulsory for replicating real-life conditions and making the result reliable for practical applications. These evaluation criteria will ensure that the neural network is free from \emph{data leakage} \cite{alneyadi2016survey} during testing and can not overperform by learning features specific to the device or scene. Besides, the performances of CMI and image manipulation detection should be measured using images manipulated at different intensities. We strictly follow the above-mentioned points in our experiments. 

In this paper, we propose a general-purpose novel CNN architecture, called L2-constrained Remnant Convolutional Neural Network (L2-constrained RemNet) for performing two crucial tasks in image forensics, CMI and image manipulation detection. Our proposed CNN has two parts, a preprocessor block and a classification block. The preprocessor architecture consists of several data-driven remnant blocks, and an L2 loss is applied to the output of the preprocessor block. A CNN based classification block follows the preprocessor block, and categorical crossentropy loss is calculated based on its output. The total loss function is a combination of the L2 loss and the categorical crossentropy loss. The whole network is trained end-to-end while minimizing the total loss. The L2-constrained preprocessor learns to suppress image contents making it easier for the classification block to extract image forensics features. Our experiments show that the proposed method can outperform other state-of-the-art networks in both image forensic tasks.

We organize the rest of the paper as follows. Section 2 contains a description of our proposed CNN and loss function. We discuss our training and evaluation criteria, along with the experimental results in section 3. Finally, we conclude in Section 4.

\begin{figure*}[!b]
	\centering
	\includegraphics[width = 4in]{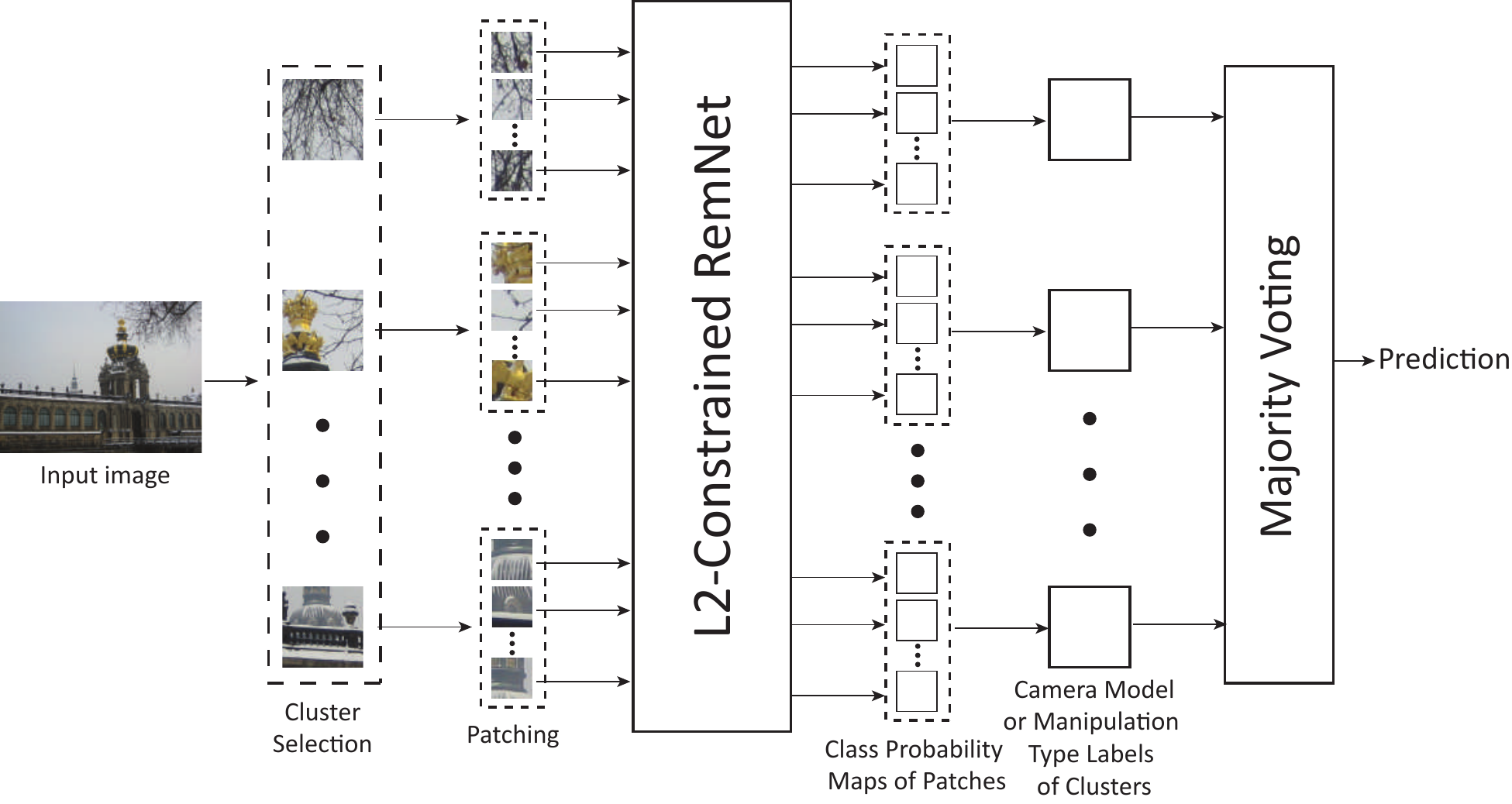}
	\caption{Schematic representation of the proposed method for CMI and image manipulation detection.}
	\label{fig_method}
\end{figure*}

\section{Proposed Method}
In this paper, we propose a CNN-based patch-level method for CMI and image manipulation detection. A schematic representation of our proposed method is shown in Fig. \ref{fig_method}.

\noindent As shown, we first extract high quality clusters of size $256\times256$ from an input image. From each cluster, patches of size $64\times64$ are taken and fed to the L2-constrained RemNet. It then generates a class probability map for each patch. We assign a camera model or image manipulation type label to each cluster by averaging the class probability maps of its patches. The final prediction is made based on the majority voting on the labels of the clusters of an image. 

\begin{figure*}[!t]
	\centering
	\includegraphics[width = 4.5in]{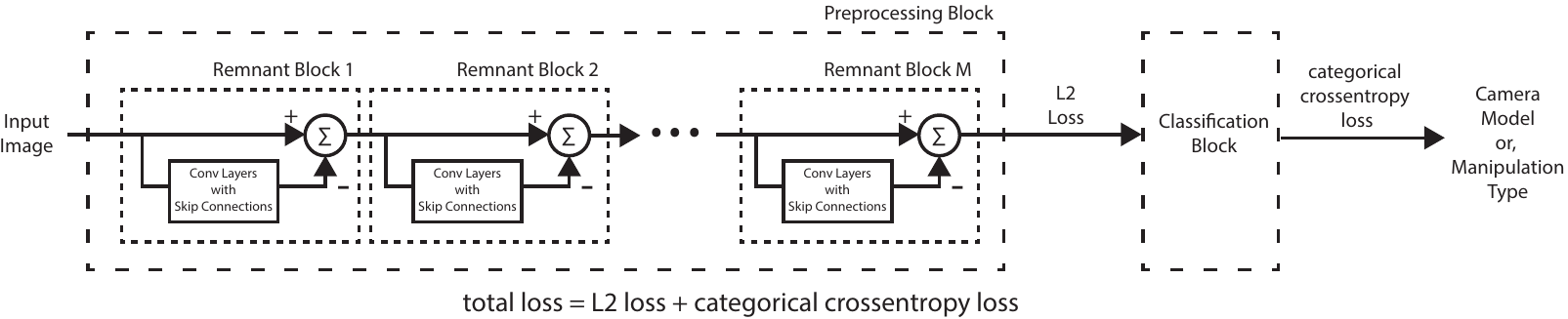}
	\caption{Line diagram of proposed modular network.}
	\label{fig_network1}
\end{figure*}

As well known, CNNs in their standard form tend to learn content-specific features from the training images. In designing CNNs for image forensic tasks, it has been, therefore, a common practice to use a preprocessing scheme to suppress the image contents and intensify the minute signatures induced by the image acquisition pipeline or image manipulation operation. Unlike the conventional approaches, the benefit of designing a dynamic preprocessing block is that it can adapt itself optimally to perform different image forensic tasks. To this end, we propose a general-purpose novel CNN architecture, called L2-constrained RemNet. A data-driven preprocessing block coupled with L2 loss is used at the beginning of the network, which is followed by a classification block. A line diagram of the proposed network model is shown in Fig. \ref{fig_network1}. The details of our proposed model are presented in the following.

\begin{figure*}[!t]
	\centering
	\includegraphics[width = 4.5in]{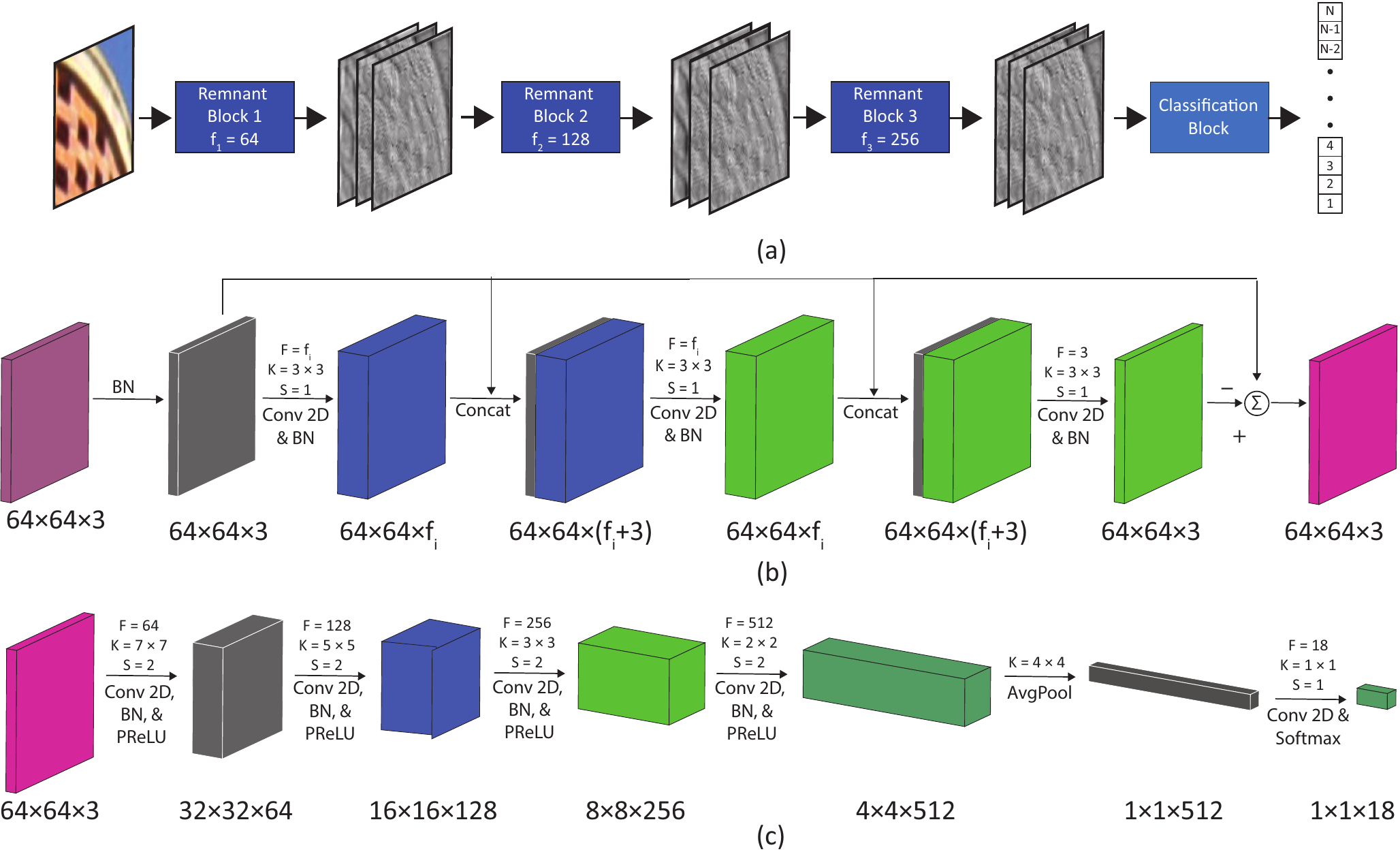}
	\caption{Architecture of our proposed L2-constrained RemNet. The overall architecture with preprocessor block and the classification block is illustrated in (a). The structure of the remnant blocks and the classification blocks are shown in (b) and (c), respectively. Here, AvgPool and Conv2D stand for average pooling and 2D convolution, respectively. The letters F, K, and S represent the number of filters, their kernel size, and strides, respectively.}
	\label{fig_network2}
\end{figure*}

\subsection{Preprocessing block}

We use the remnant blocks proposed in \cite{rafi2019remnet} as our preprocessing block. The architecture is influenced by the highway networks \cite{srivastava2015highway}. The inherent camera model-specific features are very subtle and minute features of the image \cite{stamm2013information}, \cite{lukas2006digital}, \cite{chen2008determining}. The problem of diminishing minute model-specific features is alleviated in the remnant block through the use of skip connections. Also, it refrains the minute features from being lost in a layer. Moreover, it resolves the vanishing gradient problem \cite{tong2017image} during training. The use of activation is avoided in the remnant blocks as per the design requirements in \cite{bayar2017design}. Also, it is motivated by our wish to make them perform as optimal digital filters. The final layer of a remnant block is subtracted from its input in a pixelwise manner. This subtraction helps regulate information flow. While choosing the depth of a remnant block, the number of filters in each convolutional layer, and kernel size-- we use the hyperparameters proposed in \cite{rafi2019remnet}. The architecture of the remnant block is illustrated in Fig. \ref{fig_network2}. 


Each remnant block has three convolutional layers. The kernel size is chosen as $3 \times 3$. Each layer is followed by BN (batch normalization). The feature space is widened from $64 \times 64 \times 3$ to $64 \times 64 \times f_i$ in the first two convolutional layers and then reduced to the original dimension again. The output of the last convolutional layer is subtracted from the input. As the convolutional layers are followed by BN, the input to the block is batch normalized as well. To preserve input information throughout a block, the input is propagated to every convolutional layer inside the block. We use three remnant blocks in total. 64, 128, and 256 are chosen as $f_i$ for the consecutive remnant blocks.

\subsection{Classification Block}

\begin{table}[!t]
	\centering
	\caption{Architecture of Our Proposed L2-constrained RemNet}
	\begin{tabular}{C{1in} C{1in} C{1in}}
		\hline
		Layers & Output Size & Kernels*\\
		\hline
		Remnant Block 1 & 64$\times$64$\times$3 & $f_1$ = 64\\
		
		Remnant Block 2 & 64$\times$64$\times$3 & $f_2$ = 128\\
		
		Remnant Block 3 & 64$\times$64$\times$3 & $f_3$ = 256\\
		\hline
		\multicolumn{3}{c}{\textbf{Classification Block}}\\
		\hline
		Conv 2D, BN, \& PReLU & 32$\times$32$\times$64 & F = 64, K = 7$\times$7, S = 2\\
		
		Conv 2D, BN, \& PReLU & 16$\times$16$\times$128 & F = 128, K = 5$\times$5, S = 2\\
		
		Conv 2D, BN, \& PReLU & 8$\times$8$\times$256 & F = 256, K = 3$\times$3, S = 2\\
		
		Conv 2D, BN, \& PReLU & 4$\times$4$\times$512 & F = 512, K = 2$\times$2, S = 2\\
		
		Average Pool & 1$\times$1$\times$512 & K = 4$\times$4\\
		
		Conv 2D & 1$\times$1$\times$N & F = N, K = 1$\times$1, S = 1\\
		
		Softmax & N & -- \\
		\hline
		\multicolumn{3}{p{3in}}{* \footnotesize{Here, F, K, and S represent the number of filters, kernel size, and strides, respectively. N represents the number of class.}}
	\end{tabular}
	\label{tab_arch}
\end{table}

We use the classifier proposed in \cite{rafi2019remnet} as our classification block as well. The output of the preprocessing block is passed a classification block. The architecture of the classification block is provided in Table \ref{tab_arch}. It extracts higher-level image forensics features by gradually reducing the dimensions of the feature space, and finally provide a class probability of the source camera model or the manipulation type of the input image.  

The classification block starts with four convolutional layers. Each of the first four convolutional layers is followed by a BN layer and a PReLU activation. The output of the fourth convolutional layer is followed by an average-pooling operation. Lastly, a final convolutional layer with softmax activation is used to generate a probability for the final prediction. The design choices for the classification block are the same as the hyperparameters proposed in \cite{rafi2019remnet}. There are no fully connected layers in the classification block, which keeps the number of parameters less. Consequently, the network is less prone to overfitting and trains within a shorter time. 

\subsection{Loss Function}
\label{sec_loss_function}
The preprocessing block contains $M$ remnant blocks. The $i$-th remnant block applies a transformation $H_{i}$ on its input $\mathbf{x_{i}}$ (which is also the output of the $(i-1)$-th remnant block) and subtracts it from its input to produce the output $\mathbf{y}_{\mathbf{p_{i}}}$:
\begin {equation}
\mathbf{y_{\mathbf{p_{i}}}}= \mathbf{x_{i}} - H\left(\mathbf{x_{i}}, \mathbf{W}_{\mathbf{p_{i}}}\right),
\end {equation}
The output of the last remnant block is $\mathbf{y}_{\mathbf{p_{M}}}$. A loss is calculated based on a flattened version of this output:

\begin{equation}
L_2 = \sum_{l=1}^{N_{param}} y_{p_{M_l}}^{2}.
\end{equation}
Here, $\mathbf{y}_{\mathbf{p_{M_l}}}$ is the $l$-th element of $\mathbf{y}_{\mathbf{p_{M}}}$ and $N_{param}$ is the total number of elements in $\mathbf{y}_{\mathbf{p_{M}}}$. Afterwards, $\mathbf{y}_{\mathbf{p_{M}}}$ is fed the classifier block that applies a transformation $G$ to generate the final output $\mathbf{y}_{\mathbf{c}}$:
\begin {equation}
\mathbf{y_{\mathbf{c}}}=G\left(\mathbf{y_{\mathbf{p_{M}}}}, \mathbf{W}_{\mathbf{c}}\right).
\end {equation}	
We calculate categorical crossentropy loss between this output and the ground truth using:
\begin{equation}
L_{xent} = \sum_{k=1}^{N_{class}} y_{c_{i}}^{*(k)} \log \left(y_{c_{i}}^{(k)}\right).
\end{equation}
where $y_{c_i}^{*(k)}$ and $y_{c_i}^{(k)}$ are the true label and the network output of the $i$-th image at the $k$-th class among the $N_{class}$ classes, respectively,. The total loss $L$ is defined using the following equation:

\begin{equation}
L = \alpha * L_2 + L_{xent}.
\end{equation}

Here, $\alpha$ indicates how much weight we want to put in the suppression of the residue from the preprocessor block. A larger choice for $\alpha$ may cause the vanishing gradient problem for the classifier \cite{tong2017image}. We empirically set the value of $\alpha$ as 0.5. During backpropagation, the gradient of $L_2$ is used to update the weights of the preprocessing block. The gradient of $L_{xent}$ is used to update the weights of both the preprocessing block and the classifier block. The whole network is trained in an end-to-end manner. The preprocessing block outputs a residue of the input, and $L_2$ attempts to minimize this output, which results in suppression of image contents. Simultaneously, the classifier tries to extract useful features from this residue for accurate predictions to minimize $L_{xent}$. Minimization of $L$ results in rich image forensics features in the residue for the classifier block.

\section{Experimental Results}

We perform a number of experiments to prove the efficacy of our proposed method. We discuss the experiments and the results in this section. 

\subsection{Camera Model Identification}

We evaluate our L2-constrained RemNet on Dresden Dataset \cite{dresden}. The dataset
contains images captured with 73 devices of 27 different camera models. Multiple snaps have been captured from different scenes for each device. We discard eight camera models to choose the specific camera models, which have images captured using more than one device. Our goal is to keep one device excluded during training and use it only for testing. Also, we consider Nikon D70 and Nikon D70s, as a single camera model, according to \cite{kirchner2015forensic}. We end up with 18 camera models. We split the dataset into train, validation, and test sets following strict criteria that the camera device and scenes used during testing are not used for training or validation. This results in 7938, 1353, and 540 images in the train, validation, and test set, respectively. These criteria, proposed in \cite{kirchner2015forensic} and used in \cite{bondi2017first}, is quite necessary to make sure that the evaluation is not biased owing to device-specific features and scene-specific features.
 
Data augmentation is a commonly used method in deep learning to reduce overfitting. Recently, researchers have started using it for CMI as well \cite{yang2020survey}, \cite{rafi2019application}, \cite{rafi2019remnet}. Also, our goal is to perform CMI from both unaltered and manipulated images. Therefore, we choose different image manipulations as our data augmentations, as proposed in \cite{rafi2019remnet}. The types of augmentation that we use in this work are:

\begin{itemize}
	\item[$\circ$] JPEG-Compression with a quality factor of 70\%, 80\%, and 90\%
	\item[$\circ$] Resizing by a factor of 0.5, 0.8, 1.5, and 2.0
	\item[$\circ$] Gamma-Correction with a factor of $\gamma$ = 0.8 and 1.2
\end{itemize}

This increases our data by nine folds. Afterward, we extract clusters of $256 \times 256$ size from the images. However, saturated and flat regions inside an image are not less likely to contain features related to CMI. Therefore, we follow the selection strategy proposed in \cite{bondi2017first}, \cite{rafi2019remnet} to extract high quality image clusters. For every cluster $\mathcal{P}$ in an image, its quality $Q(\mathcal{P})$ is computed as

\begin{equation}
\label{eqn_quality}
Q(\mathcal{P})=\frac{1}{3} \sum_{c\, \in[R, G, B]} \left[\alpha \cdot \beta \cdot(\mu_c - \mu_{c}^2) + (1 - \alpha) \cdot(1 - e^{\gamma \sigma_c}) \right]
\end{equation}

where $\alpha$, $\beta$, and $\gamma$ are empirically set constants (set to 0.7, 4 and $\ln (0.01)$, respectively), $\mu_c$ and $\sigma_c$, $c \in [R, G, B]$ are the mean and standard deviation of the red, green, and blue components of cluster $\mathcal{P}$, respectively.

\begin{table}[!b]
	\centering
	\caption{Accuracy (in \%) of different methods in CMI for unaltered test dataset}
	\begin{tabular}{C{1.5in} C{1.6in}}
		\hline
		Method & Dresden Dataset \\ 
		\hline
		Yang et al. \cite{yang2017source} & 95.19 \\ 

		Bayar and Stamm. \cite{bayar2017design} & 93.89 \\ 

		Bondi et al. \cite{bondi2017first} & 92.59 \\ 

		DenseNet \cite{densenet} & 95.05 \\ 

		ResNet \cite{he2016deep} & 95.19 \\ 
		
		ResNeXt \cite{xie2017aggregated} & 95.55\\
		
		RemNet without preprocessing block \cite{rafi2019remnet} & 95.74 \\ 
		
		RemNet \cite{rafi2019remnet} & 97.59 \\ 

		\textbf{Proposed Method} & \textbf{98.15} \\ 
		\hline
	\end{tabular}
	
	\label{tab_result}
\end{table}

Although we extract $256 \times 256$ sized high quality clusters, we use $64 \times 64$ input size for our network according to \cite{yang2017source}, \cite{bondi2017first}, \cite{yao2018robust}, \cite{rafi2019remnet}. Patches of $64 \times 64$ are randomly selected from a cluster of $256 \times 256$ during training. This strategy introduces statistical variations during training which is discussed in detail in \cite{rafi2019remnet}. We extract 20 clusters of size $256 \times 256$ from each image and this results in 1587600 and 270600 train and validation clusters. We use our custom loss function (see subsection 3.3) and Adam \cite{kingma2014adam} optimizer with exponential decay rate factors $\beta_1$ = 0.9 and $\beta_2$ = 0.999. The choice for our batch size is 64. The learning rate starts with $10^{-3}$ and we decrease it with a factor of $0.5$ if the softmax classification loss ($L_{xent}$) does not decrease in three successive epochs. We train our network for a maximum of 70 epochs and save the weight with the least validation softmax classification loss for evaluation. 

The test set contains $10800$ clusters of size $256 \times 256$ from $540$ full images. During testing, we average the predictions on all non-overlapping patches of size $64 \times 64$ to make a prediction for a cluster and assign a camera model label $\hat{L_{n}}$ to it. We use majority voting to make the final prediction $\hat{L}$ for the full image. The metric we use for evaluating the performance of our models is provided in the following equation:

\begin{equation}
\label{eqn_acc}
Accuracy = \frac{N_{C}}{N_{T}}.
\end{equation}

Here, $N_{C}$ is the number of correct prediction and $N_{T}$ is the total number of test images. We also compare our results with four other state-of-art CNNs in CMI \cite{bayar2017design}, fusion residual networks \cite{yang2017source}, \cite{bondi2017first}, and \cite{rafi2019remnet}. Moreover, we provide comparison with two other popular deep CNNs, ResNet \cite{he2016deep} and DenseNet \cite{densenet} as they have been used for image forensics as well \cite{Chen_densenet}, \cite{barni-jpeg-cnn}, \cite{DL_cmi}, \cite{rafi2019application}. We use the same input size for all the networks for fair comparison.

\begin{table*}[!t]
	\centering
	\caption{Accuracy (in \%) of different methods in CMI for manipulated test dataset}
	\fontsize{8.5}{10}\selectfont
	\begin{tabular}{C{0.5in} C{0.31in} C{0.31in} C{0.31in} C{0.31in} C{0.31in} C{0.31in} C{0.31in} C{0.31in} C{0.31in} C{0.31in} C{0.31in} C{0.31in}}
		\hline
		Method & \multicolumn{4}{c}{Gamma Correction} & \multicolumn{4}{c}{JPEG Compression} & \multicolumn{4}{c}{Resize Scale} \\
		\hline
		& 0.5 & 0.75 & 1.25 & 1.5 & 95 & 90 & 85 & 80 & 0.8 & 0.9 & 1.1 & 1.2 \\
		\hline
		Yang et al. \cite{yang2017source} & 94.26 & 95.37 & 95.00 & 92.78 & 94.07 & 94.07 & 92.59 & 92.59 & 94.26 & 92.59 & 90.93 & 90.56 \\
		Bayar and Stamm. \cite{bayar2017design} & 93.52 & 94.44 & 94.44 & 94.63 & 92.59 & 94.81 & 88.15 & 85.74 & 88.15 & 87.04 & 64.44 & 59.07 \\
		Bondi et al. \cite{bondi2017first} & 85.92 & 91.85 & 89.07 & 92.03 & 84.07 & 85.92 & 91.48 & 90.74 & 92.56 & 92.77 & 91.48 & 89.44 \\
		DenseNet \cite{densenet} & 91.66 & 95.18 & 92.03 & 94.62 & 92.77 & 92.96 & 94.26 & 94.81 & 95.00 & 94.81 & 94.44 & 94.26 \\
		ResNet \cite{he2016deep} & 91.85 & 95.18 & 92.77 & 94.81 & 93.88 & \textbf{94.82} & 95.55 & 95.00 & 95.18 & 95.18 & 95.00 & 95.18 \\
		ResNeXt \cite{xie2017aggregated} & 94.25 & 95.55 & 93.88 & 95.18 & 95.18 & \textbf{94.82} & 94.25 & 94.07 & 95.00 & 95.00 & \textbf{96.11} & \textbf{95.55} \\
		RemNet \cite{rafi2019remnet} & 96.11 & 97.22 & 96.11 & 95.56 & \textbf{97.59} & \textbf{94.82} & 92.59 & 92.78 & 95.00 & 93.33 & 92.04 & 92.41 \\
		Proposed Method & \textbf{96.29} & \textbf{98.14} & \textbf{97.59} & \textbf{97.96} & 92.96 & 93.33 & \textbf{96.11} & \textbf{97.03} & \textbf{96.67} & \textbf{96.67} & 90.74 & 91.66 \\
		\hline
	\end{tabular}
	
	\label{tab_manipulation}
\end{table*}

At first, we evaluate the performance of the models on the unaltered test dataset. L2-constrained RemNet achieves an overall accuracy of $98.15\%$, which is better than all other approaches we compare with (see Table \ref{tab_result}). It should be noted that we set the value for $\alpha$ in our custom loss function (5) empirically. We have achieved accuracy of 97.77\%, 98.15\%, and 97.77\%, when $\alpha$ is chosen as 0.1, 0.5, and 1, respectively. Therefore, we propose using $\alpha = 0.5$. 

We perform several experiments to justify the use of the L2-constrained preprocessing block in our network. First, we train the RemNet without any preprocessing block at the beginning of the network, that is, we only train the classification block. Then, we train the RemNet without any auxiliary L2 loss at the output of the preprocessing block. Afterward, we experiment with replacing the L2 loss with the L1 loss. The lower accuracy  of the RemNet without the  preprocessing block justifies the use of the preprocessing step. Similarly, the lower accuracy of RemNet without any additional loss justifies the use of the auxiliary loss. When we use the L1 loss in our custom loss function, the total loss oscillates throughout the training and does not converge. After a complete run, the L1-constrained RemNet attains an accuracy of 58.88\%. The L1 loss enforces sparsity on the output of the preprocessing block, whereas the image forensics features, in this case, are non-sparse and present throughout the image. The L2 loss forces the output of the preprocessing block to be small and provides a non-sparse solution.

Furthermore, we apply various manipulations on the test set and evaluate the performance of our method. To make sure that the network has not overfitted on the manipulation factors used during training, we also manipulate the test images with factors that are not used during training. The test images are created using gamma correction with $\gamma$ = 0.5, 0.75, 1.25, and 1.5; JPEG compression quality factors (QFs) 95\%, 90\%, 85\%, and 80\%; and resize scaling factor of 0.8, 0.9, 1.1, 1.2. The highest result for each manipulation factor is made bold (see Table \ref{tab_manipulation}). We can see that our proposed method has substantial improvement over other methods for Gamma Correction. In the case of JPEG Compression, our network achieves better performance for two factors, and RemNet \cite{rafi2019remnet} achieves better performance in two. For Resize manipulation, we see that ResNeXt \cite{xie2017aggregated} gains higher accuracy for two manipulation factors, whereas our proposed method gains higher accuracy in the other two factors. We can conclude that our proposed method proves to be most robust to external manipulation. Also, deep CNNs perform better than shallow networks in the face of manipulated images.

\subsection{Image Manipulation Detection}

Now, we show the use of our network in a completely different image forensic task. We use it to identify the kind of image-manipulation done on an image. The same network is used here except the number of output classes, which is four-- unaltered, rescale, JPEG compression, and gamma correction. The input size for all the networks is also maintained at ($64 \times 64$). We use the same train and validation set from our experiments with CMI and sub-divide it into the four manipulation classes. The L2-constrained RemNet is then trained to detect the type of manipulation applied to an image. It is to be mentioned that, during training, our dataset consisting of 1587600 train and 270600 validation clusters has been reduced in order to make the training data evenly distributed among four classes. Since the number of unaltered train and validation clusters are 158760 and 27060, respectively, we select 158760 train and 27060 validation clusters randomly for each type of manipulation.

\begin{table}[!b]
	\centering
	\caption{Accuracy (in \%) of different methods in image manipulation detection}
	\begin{tabular}{C{1.5in} C{1.6in}}
		\hline
		Method & Dresden Dataset \\ 
		\hline
		Yang et al. \cite{yang2017source} & 91.74 \\ 
		
		Bayar and Stamm \cite{bayar2017design} & 87.28 \\ 
		
		RemNet \cite{rafi2019remnet} & 98.27 \\ 
		
		\textbf{Proposed Method} & \textbf{99.68} \\ 
		\hline
	\end{tabular}
	
	\label{tab_result_manip}
\end{table}
\begin{table*}[!t]
	\centering
	\fontsize{8.5}{10}\selectfont
	\caption{Accuracy (in \%) of image manipulation detection for different manipulation factors}
	
	\begin{tabular}{C{0.5in} C{0.34in} C{0.34in} C{0.34in} C{0.34in} C{0.34in} C{0.34in} C{0.34in} C{0.34in} C{0.34in} C{0.34in} C{0.34in}}
		\hline
		Method & \multicolumn{4}{c}{Gamma Correction} & \multicolumn{3}{c}{JPEG Compression} & \multicolumn{4}{c}{Rescale} \\
		\hline
		& 0.5 & 0.75 & 1.25 & 1.5 & 95 & 90 & 85 & 0.8 & 0.9 & 1.1 & 1.2 \\
		\hline
		Yang et al. \cite{yang2017source} & 99.07 & 98.52 & 97.04 & 98.70 & 49.44 & 100 & 100 & 100 & 97.40 & 60.74 & 100 \\
		
		Bayar and Stamm \cite{bayar2017design} & 94.44 & 83.33 & 77.22 & 90.56 & 11.30 & 100 & 100 & 100 & 100 & 90.93 & 99.63 \\
		RemNet \cite{rafi2019remnet} & 100 & \textbf{99.81} & \textbf{99.63} & \textbf{100} & 81.48 & 98.33 & 100 & 100 & 100 & 100 & 100 \\
		Proposed method & \textbf{100} & 99.63 & 99.26 & 98.7 & \textbf{100} & \textbf{98.7} & \textbf{100} & \textbf{100} & \textbf{100} & \textbf{100} & \textbf{100} \\
		\hline
	\end{tabular}
	\label{tab_manip_detect}
\end{table*}

In testing, we have used the test images from the Dresden dataset and generated a total of $540 \times 12 = 6480$ test images, which include $540$ unaltered images; $540 \times 4 = 2160$ gamma-corrected images with $\gamma$ = 0.5, 0.75, 1.25, and 1.5; $540 \times 3 = 1620$ JPEG compressed images compressed with factors of 85\%, 90\%, and 95\%; and $540\times4 = 2160$ resized images images with scaling factor of 0.8, 0.9, 1.1, and 1.2. Details of the results are given in Table \ref{tab_manip_detect}. We achieve an overall accuracy of 99.68\% in this task whereas RemNet \cite{rafi2019remnet}, Bayar and Stamm \cite{bayar2017design}, and Yang et al. \cite{yang2017source} achieve 98.27\%, 87.28\% and 91.74\%, respectively (see Table \ref{tab_result_manip}). We demonstrate the detection accuracy for different factors of manipulation in Table \ref{tab_manip_detect}. For gamma-corrected images, the performances of \cite{yang2017source}, RemNet \cite{rafi2019remnet} and our proposed method are substantially better than that of \cite{bayar2017design}. In the case of JPEG compression, all four networks perform almost the same except at the compression factor of 95, where \cite{bayar2017design} and \cite{yang2017source} fail miserably by misclassifying most of the compressed images as unaltered images. There is a significant drop in the detection accuracy for RemNet \cite{rafi2019remnet} as well. This is expected since there is very little difference between the original image and JPEG compressed image with factor 95. However, our proposed method achieves 100\% accuracy even at this factor, which indicates that the network can detect even minute manipulation artifacts introduced during manipulation operation. When detecting rescaled images, our network and RemNet \cite{rafi2019remnet} performs the same by attaining a 100\% accuracy. Of the other two networks, \cite{bayar2017design} performs better than \cite{yang2017source}.

\section{Conclusion}

In this paper, we have proposed an L2 loss constrained RemNet for performing two important image forensics tasks, namely, CMI and image manipulation detection. The proposed modular CNN model comprises of a dynamic preprocessor and a classification block in series. The L2-constrained preprocessor, while trained end-to-end along with the classification block, suppresses unnecessary image contents dynamically and generates a residue of the image from where the classification block can easily extract image forensics features. We have comprehensively conducted multiple experiments on the Dresden dataset to demonstrate the efficacy of such a preprocessing scheme assisted by the L2 loss in CMI. During testing, we use images captured by devices not seen during training to replicate practical applications. The results of the experiments have shown that our proposed method can be successfully used in real-world scenarios. Additionally, we have used our proposed method for image manipulation detection. The satisfactory performances of our network on both classification tasks prove that it can be used for a general-purpose network for image forensics.

%
%
\bibliographystyle{splncs04}
\bibliography{ref}
\end{document}